\documentclass{PoS}
\usepackage{amsmath}
\usepackage{amssymb}
\usepackage{epsfig}

\usepackage{fontenc}
\usepackage{times}
\usepackage{mathptmx}
\usepackage{graphicx}
\usepackage{subfigure}

\newcommand{\be}{\begin{equation}}
\newcommand{\ee}{\end{equation}}

\newcommand{\startitem}{\begin{itemize}}
\newcommand{\stopitem}{\end{itemize}}
\newcommand{\bs}{\begin{split}}
\newcommand{\es}{\end{split}}

\newcommand{\KET}[1]{\ensuremath{|#1\rangle}}
\newcommand{\BRA}[1]{\ensuremath{\langle #1|}}
\newcommand{\VEC}[1]{\ensuremath{\mbox{\boldmath{$#1$}}}}

\newcommand{\bea}{\begin{eqnarray}}
\newcommand{\eea}{\end{eqnarray}}

\title{$p\bar{p} \, \rightarrow \, \Lambda_c \bar{\Lambda}_c$
within the generalized parton picture - \\first results}

\ShortTitle{$p\bar{p} \, \rightarrow \, \Lambda_c \bar{\Lambda}_c$
within the generalized parton picture}

\author{\speaker{A.T.~Goritschnig} \\
        Institut f\"ur Physik, Karl-Franzens-Universit\"at Graz, Austria\\
        E-mail: \email{alexander.goritschnig@uni-graz.at}}

\author{P.~Kroll \\
        Fachbereich Physik, Universit\"at Wuppertal, Germany\\
        E-mail: \email{kroll@theorie.physik.uni-wuppertal.de}}

\author{W.~Schweiger \\
        Institut f\"ur Physik, Karl-Franzens-Universit\"at Graz, Austria\\
        E-mail: \email{wolfgang.schweiger@uni-graz.at}}

\abstract{ We study proton-antiproton annihilation into $\Lambda_c
\bar{\Lambda}_c$ pairs within the generalized parton picture. Our
starting point is the double handbag diagram which is shown to
factorize into soft generalized parton distributions for the $p
\rightarrow \Lambda_c$ (and $\bar{p} \rightarrow \bar{\Lambda}_c$)
transition and a hard subprocess amplitude for $u \bar{u}
\rightarrow c \bar{c}$. Thereby the mass of the charm quark is
taken as the hard scale so that our results are not restricted to
large scattering angles and/or incredibly large energies.
Modelling the generalized parton distributions for the
$p\rightarrow \Lambda_c$ transition by an overlap of simple
quark-diquark light-cone wave functions we make first predictions
for $p\rightarrow  \Lambda_c$ transition form factors and
unpolarized $p\bar{p} \rightarrow \Lambda_c \bar{\Lambda}_c$ cross
sections. Our findings may become interesting in view of
forthcoming experiments at FAIR in Darmstadt.}

\FullConference{LIGHT CONE 2008 Relativistic Nuclear and Particle Physics\\
                 July 7-11, 2008\\
                 Mulhouse, France}

\begin{document}

\section{Introduction}
Exclusive hadronic reactions which require the production of heavy
quark-antiquark pairs are, to a large extent, cleaner and easier
to handle than those in which only light flavors are involved. On
the one hand, certain elementary reaction mechanisms can already
be ruled out from the beginning due to the small heavy-flavor
content of the quark sea. On the other hand, the mass of the heavy
quark itself can serve as a hard scale so that there is a good
chance that QCD perturbation theory provides a substantial part of
the process amplitude already at moderately large energies.

\begin{center}
\begin{figure}[h]
\begin{center}
\epsfig{file=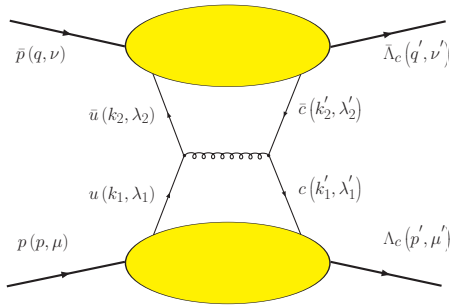,width=6cm,clip=} \caption{Double-handbag
contribution to $p\bar{p} \rightarrow \Lambda_c\bar{\Lambda}_c$.}
\label{Fig:Factorization}
\end{center}
\end{figure}
\end{center}

The simplest elementary reaction mechanism for the process we are
interested in, namely $p \bar{p} \rightarrow \Lambda_c
\bar{\Lambda}_c$, is depicted in Fig.~\ref{Fig:Factorization}.
This is the mechanism which we assume to be dominant in the
forward hemisphere for energies well above production threshold.
We are going to analyze it in terms of generalized parton
distributions~\cite{Radyushkin:1997ki},\cite{Ji:1996nm}.

\section{Kinematics}
We denote particle momenta and helicities as shown in
Fig.~\ref{Fig:Factorization}. For our purposes it is most
convenient to work in a center-of-mass frame, in which the
light-cone (LC) components of the proton and $\Lambda_c$ momenta
are parameterized as
\footnote{We use the following convention for the LC components of
a 4-vector $v$: $v = \left[ v^+ , v^- , \VEC{v}_\perp \right]$
with $v^\pm = 1/\sqrt{2} \left( v^0 \pm v^3 \right)$ and
$\VEC{v}_\perp = \left(v^1 , v^2\right)$.}
\be
  p = \left[ \left(1+\xi\right) \bar{p}^{+},
  \frac{m^2+\VEC{\Delta}_\perp^2 / 4}{2\left(1+\xi\right)\bar{p}^{+}},
  -\frac{\VEC{\Delta}_{\perp}}{2} \right]
  \, \quad \hbox{and} \quad  p^{'} = \left[ \left( 1-\xi\right) \bar{p}^{+},
  \frac{M^2 +
  \VEC{\Delta}_{\perp}^2 / 4}{2\left( 1-\xi\right)\bar{p}^{+}},
  \frac{\VEC{\Delta}_{\perp}}{2} \right],
\ee
respectively. $M$ stands for the $\Lambda_c$ mass and m for the
mass of the proton. The corresponding antiparticle momenta are $q
= \left[ p^- , p^+ , \VEC{\Delta}_{\perp}/2 \right]$ and
$q^{\prime} = \left[ p^{\prime -} , p^{\prime +} ,
-\VEC{\Delta}_{\perp}/2 \right]$. The average of proton and
$\Lambda_c$ momenta $\bar{p} = \frac12 \left(p +
p^{\prime}\right)$ defines the longitudinal direction and the
4-momentum transfer is specified by $\Delta = p^{\prime} - p = q -
q^{\prime}$. The relative momentum transfer in longitudinal
direction is given by the \lq\lq skewness parameter\rq\rq
\be \xi \equiv \frac{p^+ - p^{\prime +}}{p^+ + p^{\prime +}} = -
\frac{\Delta^+}{2\bar{p}^{+}}.
\ee

\section{Factorization}
%\label{Factorization}
%
If the dynamical mechanism underlying $p  \bar{p} \, \rightarrow
\, \Lambda_c \bar{\Lambda}_c$ scattering is given by
Fig.~\ref{Fig:Factorization}, the corresponding scattering
amplitude can be written as
($k_i^{\mathrm{av}}=(k_i+k_i^\prime)/2$)
\be
\begin{split}
& M_{\mu^{'}\nu^{'}, \mu \nu} =
 \int d^4 k_1^{\mathrm{av}} \theta\left(k_1^{\mathrm{av}\, +}\right)
 \int \frac{d^4 z_1}
 {\left(2\pi\right)^4}\,\, e^{i\, k_1^{\mathrm{av}} z_1}
 \int d^4 k_2^{\mathrm{av}} \theta\left(k_2^{\mathrm{av}\,-}\right)
 \int \frac{d^4 z_2}
 {\left(2\pi\right)^4}\,\, e^{i\, k_2^{\mathrm{av}} z_2}
 \, H \left(k_1^\prime , k_2^\prime; k_1 , k_2\right)  \\
& \times \langle \Lambda_c : p^{'}, \mu^{'} |
 T \bar{\Psi}^{c}\left(-\frac{z_1}{2}\right) \Psi^{u}\left(\frac{z_1}{2}\right)
 |  p : p, \mu \rangle
  \langle \bar{\Lambda}_c : q^{'}, \nu^{'} |
 T \bar{\Psi}^{u}\left(\frac{z_2}{2}\right) \Psi^{c}\left(-\frac{z_2}{2}\right)
 | \bar{p} : q, \nu \rangle ,
\end{split}
\label{Eq_ProcessAmplitude}
\ee
with  $H \left(k_1^\prime , k_2^\prime; k_1 , k_2\right)$
representing the scattering amplitude for the elementary
subprocess
\be u\left(k_1 , \lambda_1 \right) \, \bar{u}\left(k_2 , \lambda_2
\right) \, \rightarrow
   \, c(k_1^{'} , \lambda_1^{'} ) \, \bar{c}(k_2^{'} , \lambda_2^{'}
   )\, .
\ee
The hadronic matrix elements describe the emission of a light
(anti)quark by the (anti)proton and the absorption of a charm
(anti)quark by the (anti)$\Lambda_c$. For better readability we
have suppressed helicity labels as well as color and spinor
indices (and corresponding sums) for the quarks.

In order to simplify the right-hand-side of
Eq.~(\ref{Eq_ProcessAmplitude}) we make a few plausible
assumptions. In analogy to Compton scattering~\cite{Diehl:1998kh},
\cite{Diehl:1999tr} we assume that the process is dominated by
partons with restricted virtualities and transverse momenta. To be
more precise, we demand for the active (anti)quarks that $k_i^2
\lesssim \Lambda^2$ and $\vert k_i^{\prime\, 2} - m_c^2\vert
\lesssim \Lambda^2$ (cf. Fig.~\ref{Fig:Factorization} for the
assignement of (anti)quark momenta). The relative transverse
momentum components should satisfy
\footnote{The plus-components of parton momentum $k$ and hadron
momentum $p$ are related by $k^+ = x p^+$. For further purposes it
is also convenient to introduce an \lq\lq average\rq\rq\ momentum
fraction of the active quarks, i.e. $\bar{x}_1 =
(k_1^{\mathrm{av}\, +})/(\bar{p}^+)$ (and analogously for the
active antiquarks)}
\be \tilde{\VEC{k}}_{\perp i}^2 / x_i \, \lesssim \, \Lambda^2 ,
\quad \hat{\VEC{k}}_{\perp i}^{\prime 2} / x_i^{\, \prime} \,
\lesssim \, \Lambda^2\, ,
\ee
with $\Lambda$ being a hadronic scale of the order of $1$~GeV and
$m_c$ the charm-quark mass. The restrictions  for the transverse
momenta are supposed to hold in the hadron in/out frames
(tilde/hat), in which the corresponding parent hadrons carry no
transverse momenta. Analogous constraints should also hold for
spectator partons. Furthermore, the hadronic matrix elements
represented by the blobs in Fig.~\ref{Fig:Factorization} are
assumed to exhibit a pronounced peak at $x_0 = m_c / M \approx 0.6
- 0.7$. This assumption reflects the expected behavior of
light-cone wave functions for the $\Lambda_c$ and is also
supported by similar observations for heavy-quark fragmentation
functions.

Since the virtuality of the gluon in the subprocess amplitude has
to be at least $4 m_c^2$, it occurs to be natural to take the
charm-quark mass $m_c$ as a hard scale. Under the foregoing
assumptions on the parton momenta it is then a good approximation
to neglect the relative transverse momenta
($\tilde{\VEC{k}}_{\perp i}$  and $\hat{\VEC{k}}_{\perp
i}^{\prime}$) of the (anti)quarks in the subprocess amplitude $H
\left(k_1^\prime , k_2^\prime; k_1 , k_2\right)$ and to replace
the momenta of the active quarks $k_i^{(\prime)}$ by (on-shell)
momenta $\bar{k}_i^{(\prime)}$ which are collinear to those of the
corresponding hadrons. With these approximations in the subprocess
amplitude most of the integrations in
Eq.~(\ref{Eq_ProcessAmplitude}) can be done analytically. The
resulting delta functions enforce a light-like separation of the
fields in the hadronic matrix elements. This allows us to drop the
time ordering of the fields. Due to our assumption on the hadronic
matrix elements it is also justified to apply a \lq\lq peaking
approximation\rq\rq\ to the subprocess amplitude (i.e. replace all
$x_i$ by $x_0$). In this way we are able to write the hadronic
amplitude as a product of integrals over soft hadronic matrix
elements with a hard partonic subprocess amplitude:
\be
\begin{split}
M_{\mu^{'}\nu^{'}, \mu\nu}(p^\prime, q^\prime; p, q)  &= H
\left(\bar{k}_1^\prime , \bar{k}_2^\prime; \bar{k}_1 ,
\bar{k}_2\right)\vert_{x_i^{(\prime)}=x_0}\,\\  &\times
  \int d k_1^{\mathrm{av}\, +} \, \theta(k_1^{\mathrm{av}\, +})\,
  \int \frac{d z_1^-}{2\pi} e^{i k_1^{\mathrm{av}\, +} z_1^-}
  \langle \Lambda_c : p^{'}, \mu^{'} |
  \bar{\Psi}^{c} \left(-\frac{z_1^-}{2}\right)
  \Psi^{u} \left(\frac{z_1^-}{2}\right)
 |  p : p, \mu \rangle\\
  &\times  \int d k_2^{\mathrm{av}\, -} \theta(k_2^{\mathrm{av}\, -}) \,
 \int \frac{d z_2^+}{2\pi} e^{i k_2^{\mathrm{av}\, -} z_2^+}
  \langle \bar{\Lambda}_c : q^{'}, \nu^{'} |
  \bar{\Psi}^{u} \left(\frac{z_2^+}{2}\right) \Psi^{c} \left(-\frac{z_2^+}{2}\right)
 |\bar{p} :  q, \nu \rangle .
\end{split}
\label{Eq_ProcessAmplitudeFactorized}
\ee
The subprocess amplitude can be calculated perturbatively, whereas
the hadronic matrix elements describe the non-perturbative
transitions of the (anti-)proton to the (anti-)$\Lambda_c$ by
emission of a (anti)$u$ quark and reabsorption of a (anti)$c$
quark. It are just the hadronic matrix elements which give rise to
generalized parton distribution functions (GPDs) and to form
factors, which are essentially $1/x$ moments of the
GPDs~\cite{Radyushkin:1997ki}, \cite{Ji:1996nm}. Therefore we will
have a closer look on them.

After some algebraic manipulations (insertion of energy- and
helicity-projectors, etc.) the first double integral in
Eq.~(\ref{Eq_ProcessAmplitudeFactorized}) can be split up into
three (independent) terms of the form
\be
 \int_\xi^1 \frac{d\bar{x}_1}{\sqrt{\bar{x}_1^2-\xi^2}} \phantom{x} \bar{p}^+
 \int \frac{d z_1^-}{2\pi} e^{i \bar{x}_1 \bar{p}^+ z_1^-}
 \BRA{\Lambda_c : p^{'} \mu^{'}}\,
 \bar{\Psi}^c\, \Gamma \, \Psi^u
 \,\KET{p : p \mu } =:
 \int_\xi^1 \frac{d\bar{x}_1}{\sqrt{\bar{x}_1^2-\xi^2}} \phantom{x}
 \mathcal{H}^{cu}_{\mu^{'}\mu}\left(\Gamma\right),
\label{Eq_HadronicMatrixElements}
\ee
where $\Gamma = \gamma^+ , \gamma^{+} \gamma_5$, or
$i\sigma^{+j}$. $\gamma^+$ and $\gamma^{+} \gamma_5$ demand for
the same helicity of the emitted and absorbed parton, whereas $i
\sigma^{+j}$ is associated with a helicity flip. At this point the
GPDs are introduced by decomposing
$\mathcal{H}^{cu}_{\mu^{'}\mu}\left(\Gamma\right)$ into its
covariant structures:
\be
 \mathcal{H}^{cu}_{\mu^{'}\mu}(\gamma^+)
 = \bar{u}\left(p^{'} , \mu^{'}\right) \left[
 H^{cu}\left(\bar{x}, \xi , t\right) \gamma^+ +
 E^{cu}\left(\bar{x}, \xi , t\right) \frac{i\sigma^{+\nu}\Delta_\nu}{M+m}
 \right] u\left(p , \mu\right)\, ,
\label{Eq_MathcalH}
\ee
\be
\tilde{\mathcal{H}}^{cu}_{\mu^{'}\mu}(\gamma^+ \gamma_5)
 = \bar{u}\left(p^{'} , \mu^{'}\right) \left[
 \tilde{H}^{cu}\left(\bar{x}, \xi , t\right) \gamma^+ \gamma_5
 + \tilde{E}^{cu} \left(\bar{x}, \xi , t\right) \frac{\gamma_5
 \Delta^+}{M+m} \right] u\left(p , \mu\right)\, ,
\label{Eq_MathcalHTilde}
\ee
\begin{equation}
\begin{split}
 \mathcal{H}^{Tcu}_{j \mu^{'}\mu}(i \sigma^{+j})
 & = \bar{u}\left(p^{'} , \mu^{'}\right) \left[
 H_T^{cu} \left(\bar{x}, \xi , t\right) i \sigma^{+j} +
 \tilde{H}_T^{cu} \left(\bar{x}, \xi , t\right) \frac{\bar{p}^+ \Delta^j
 - \Delta^+ \bar{p}^j}{Mm} + \right. \\
 & \left. E_T^{cu} \left(\bar{x}, \xi , t\right)
 \frac{\gamma^+ \Delta^j - \Delta^+ \gamma^j}{M+m} +
 \tilde{E}_T^{cu} \left(\bar{x}, \xi , t\right)
 \frac{\gamma^+ \bar{p}^j - \bar{p}^+ \gamma^j}{(M+m)/2}
 \right] u\left(p , \mu\right).
\end{split}
\label{Eq_MathcalHT}
\end{equation}
Analogous results hold for the hadronic matrix elements of the
antiparticles.

What enters the scattering amplitude are rather moments of the
GPDs than the GPDs themselves. It is thus convenient to introduce
form factors
\be R_i\left( \xi , t \right) = \int_\xi^1 \,
\frac{d\bar{x}_1}{\sqrt{\bar{x}_1^2-\xi^2}} \, F_i\left( \bar{x} ,
\xi , t \right), \label{Eq_FormFactorDef} \ee
where $F_i$ stands for any of the transition GPDs introduced in
Eqs.~(\ref{Eq_MathcalH}) -- (\ref{Eq_MathcalHT}). In our notation
the form factors $R_V ,\, R_T ,\, R_A ,\, R_P ,\, S_T ,\, S_{V1}
,\, S_S ,\, S_{V2}$ correspond to the GPDs $H ,\, E ,\, \Tilde{H}
,\, \Tilde{E} ,\, H_T ,\, E_T ,\, \Tilde{H}_T ,\, \Tilde{E}_T$,
respectively.

Combining Eqs.~(\ref{Eq_HadronicMatrixElements}) --
(\ref{Eq_FormFactorDef}) and inserting them into
Eq.~(\ref{Eq_ProcessAmplitudeFactorized}) allows us to write the
hadronic scattering amplitude in terms of these, a priori unknown,
form factors and the hard subprocess amplitude. The result is a
rather lengthy expression. Fortunately, some of the form factors
can be neglected. By taking different helicity combinations for
the matrix elements in Eqs.~(\ref{Eq_MathcalH}) --
(\ref{Eq_MathcalHT}) we observe that all the form factors apart of
$R_V , R_A$ and $S_T$ (corresponding to the GPDs $H , \Tilde{H}$
and $H_T$) must vanish in forward direction, since they are
connected with orbital angular momentum. Also for large Mandelstam
$t$ these form factors should be suppressed (as compared to $R_V ,
R_A$ and $S_T$) by powers of $\sqrt{-t}$, depending on how many
units of angular momentum are involved. We therefore expect the
following hierarchy of the form factors:
\be |R_{V}| \, , |R_{A}| ,\,  |S_{T}| \, >> \, |R_{T}| ,\,
|R_{P}| ,\, |S_{S}| ,\, |S_{V1}| ,\,  |S_{V2}|.
\label{Eq_FFHierarchy} \ee
In $M_{\mu^{'}\nu^{'}, \mu\nu}(p^\prime, q^\prime; p, q)$ a
further suppression of contributions coming from helicity flip
form factors is caused by the corresponding subprocess amplitudes.
We therefore neglect all helicity-flip form factors apart of
$S_T$. This provides us with comparably simple expressions for the
$p \bar{p} \, \rightarrow \, \Lambda_c \bar{\Lambda}_c$ helicity
amplitudes in terms of the form factors $R_V , R_A$ and $S_T$ and
the subprocess amplitudes:
\begin{eqnarray}
M_{+\pm , +\pm} &=& \frac{C_F}{2N_C} \left(1-\xi^2\right) H_{+-,+-} \left( R_V^2 \mp R_A^2 \right), \\
M_{-- , ++} &=& 0, \\
M_{-+ , +-} &=& \frac{C_F}{N_C} \left(1-\xi^2\right) H_{-+,+-} S_T^2 , \\
M_{++ , \pm\mp} &=& \frac{C_F}{2N_C} \left(1-\xi^2\right) H_{++,+-} S_T \left( R_V + R_A \right), \\
M_{\mp\pm , ++} &=& \mp \frac{C_F}{2N_C} \left(1-\xi^2\right) H_{++,+-} S_T \left( R_V - R_A \right),
\label{Eq_HelicityAmplitudes}
\end{eqnarray}
with
\begin{eqnarray}
H_{+-,+-} &=& -8\pi\, \alpha_s (x_0^2 s )\,  \cos^2
(\theta_{\mathrm{cm}} / 2 )\, , \qquad H_{-+,+-} = 8\pi\, \alpha_s
(x_0^2 s )\,  \sin^2 (\theta_{\mathrm{cm}} / 2)\, , \\ H_{++,+-}
&=& -8\pi\, \alpha_s (x_0^2 s )\,  \frac{M}{\sqrt{s}}\,
\sin\theta_{\mathrm{cm}}\, , \qquad\quad\!  H_{++,-+} =
H_{++,+-}\, .
\end{eqnarray}
$C_F=4/3$ denotes the color factor, $N_C=3$ is the
number of colors. The remaining helicity amplitudes are
related by parity invariance,
\be M_{-\mu^{\prime}-\nu^{\prime} , -\mu -\nu} =
\left(-1\right)^{\mu^{\prime}-\nu^{\prime}-\mu+\nu} \,
M_{\mu^{'}\nu^{'} , \mu\nu}\, .
\ee
For a more detailed account of these results and their derivation
 we refer to Ref.~\cite{Goritschnig:2008}.

\section{Modelling the GPDs}
The kinematical requirement for the production of a $c\bar{c}$
pair $x \gtrsim 2 x_0 M/\sqrt{s}$ implies that $x >
\xi$.\footnote{This is also the reason why the integral in
Eq.~(\ref{Eq_FormFactorDef}) starts from $\xi$ and not from $0$.}
We are thus in the DGLAP region where GPDs can be modelled as
overlap of light-cone wave functions of the proton and the
$\Lambda_c$~\cite{Diehl:2000xz}. For the $\Lambda_c$ it is
certainly a good approximation to consider only its valence Fock
state. For the proton higher Fock states may be important,
but the required overlap with the $\Lambda_c$ projects out only
appropriate spin-flavor combinations of its valence Fock state.

As a first attempt we consider a simple quark-diquark model for
the baryons. We assume that a baryon consists of an active quark,
which undergoes the hard scattering, and a diquark, which acts as
a spectator. For the $\Lambda_c$ the diquark has to be a
spin-isospin scalar. In Ref.~\cite{Kroll:1988cd} $p \bar{p} \,
\rightarrow \, \Lambda_c \bar{\Lambda}_c$ has already been studied
within a quark-diquark model, but without using the general
framework of GPDs. For our present investigation we will take a
quark-diquark wave function for the $\Lambda_c$  which is similar
to the one used for a calculation of heavy-baryon transition form
factors~\cite{Korner:1992uw}:
\be \KET{\Lambda_c^+ : \pm} = \KET{c_{\pm} S_{\left[ud\right]}}
\quad \text{with} \quad \Psi_{\Lambda}(x,k_\perp)  = N_{\Lambda}
\, \left(1-x\right) \,
e^{-\frac{a_{\Lambda}^2}{x\left(1-x\right)}\VEC{k}_{\perp}^2} \,
e^{ - a_{\Lambda}^2 M^2
\frac{\left(x-x_0\right)^2}{x\left(1-x\right)}}.
\label{Eq_LambdaWaveFct} \ee
$x$ denotes the momentum fraction carried by the quark. The
quark-diquark wave function of the proton, on the other hand, is
chosen similar to the one in Ref.~\cite{Kroll:1988cd}:
\be \KET{p: \pm}(x,k_\perp) = \KET{u_{\pm} S_{\left[ud\right]}}
\quad \text{with} \quad \Psi_p(x,k_\perp)  = N_p \,
\left(1-x\right) \, e^{-
\frac{a_p^2}{x\left(1-x\right)}\VEC{k}_{\perp}^2}.
\label{Eq_ProtonWaveFct} \ee
These are pure $s$-wave wave functions and hence all GPDs apart of
$H$, $\tilde{H}$ and $H_T$ will vanish. Within this model the
quark should have the same helicity as the baryon, which implies
$\Tilde{H}=H$. Likewise, also $H_T$ is non-zero and we even have
\be H_T = \Tilde{H} = H \, ,
\ee
i.e. we are left with a single GPD. For the model wave functions
(\ref{Eq_LambdaWaveFct}) and (\ref{Eq_ProtonWaveFct}) the overlap
integral can be done analytically  with the result
\be
\begin{split}
H&(\bar{x},\xi,\Delta_\perp^2)  =  \left(\frac{N_{\Lambda}
N_p}{16\pi^2}\right) \frac{1}{\left(1-\xi\right)^{3/2}}
\frac{\left(1-\bar{x}\right)^3 \left(\bar{x}^2 -
\xi^2\right)}{a_{\Lambda}^2 \left(1-\xi\right)^2
\left(\bar{x}+\xi\right) + a_p^2 \left(1+\xi\right)^2
\left(\bar{x}-\xi\right)} \\
& \times \exp\left[ - a_{\Lambda}^2 M^2 \frac{\left(\bar{x} - \xi
- x_0\left(1-\xi\right)\right)^2} {\left(\bar{x}-\xi\right)
\left(1-\bar{x}\right)}\right]
 \exp\left[- \VEC{\Delta}_{\perp}^2 \frac{a_{\Lambda}^2 a_p^2 \left(1-\bar{x}\right)}
 {a_{\Lambda}^2 \left(1-\xi\right)^2
\left(\bar{x}+\xi\right) + a_p^2
\left(1+\xi\right)^2\left(\bar{x}-\xi\right)}\right].
\end{split}
\label{Eq_GPD}
\ee
With this GPD we can now evaluate the form factor $R \equiv R_V =
R_A = S_T$.

\section{Numerical Results and Conclusions}
\begin{figure}[hbt]
   \subfigure{\epsfig{file=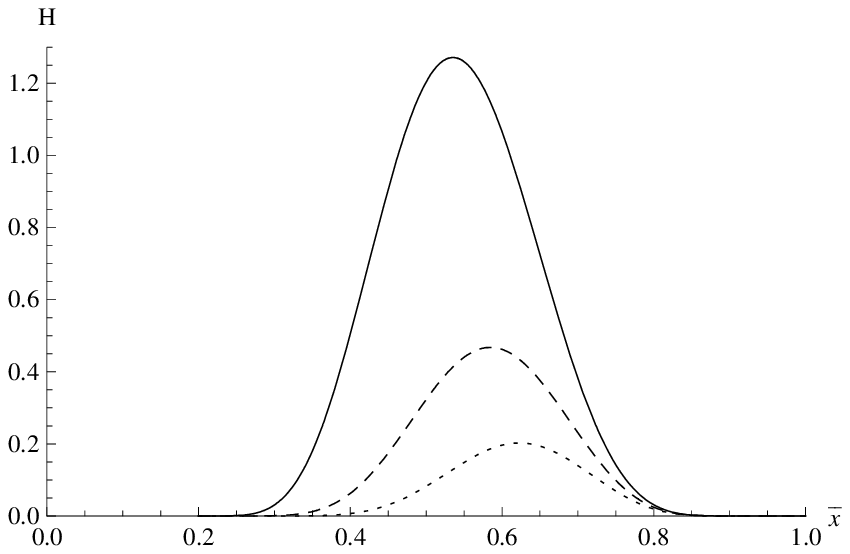,height=4.3cm,clip=}}
   \hfill
   \subfigure{\epsfig{file=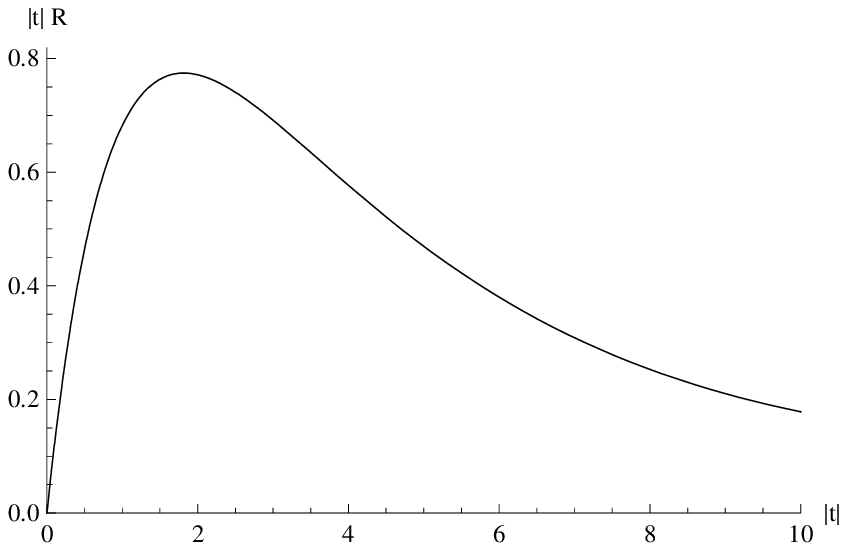,height=4.3cm,clip=}}
   \vspace{-0.3cm}\caption{{\em Left:} The transition GPD $H$ in the quark-diquark model versus $\bar{x}$ at
   Mandelstam $s = 30~\text{ GeV}^2$ for $\Delta_{\perp}^2 = 0 \text{ (solid)}, \, 2
   \text{ (dashed)} , \, 4 \text{ (dotted)} \text{ GeV}^2$
   (corresponding to $-t = 1.13 ,\, 3.30 ,\, 5.54 \text{GeV}^2$ or $\xi = 0.11 ,\,
   0.12 ,\, 0.14$).
   {\em Right:} The transition form factor $R$ scaled by $|t|$ in the quark-diquark
   model versus $|t| \, \left[\text{GeV}^2\right]$ for Mandelstam $s=30\text{ GeV}^2$.}
    \label{Fig_GPDandFF}
\end{figure}

\begin{figure}[hbt]
   \subfigure{\epsfig{file=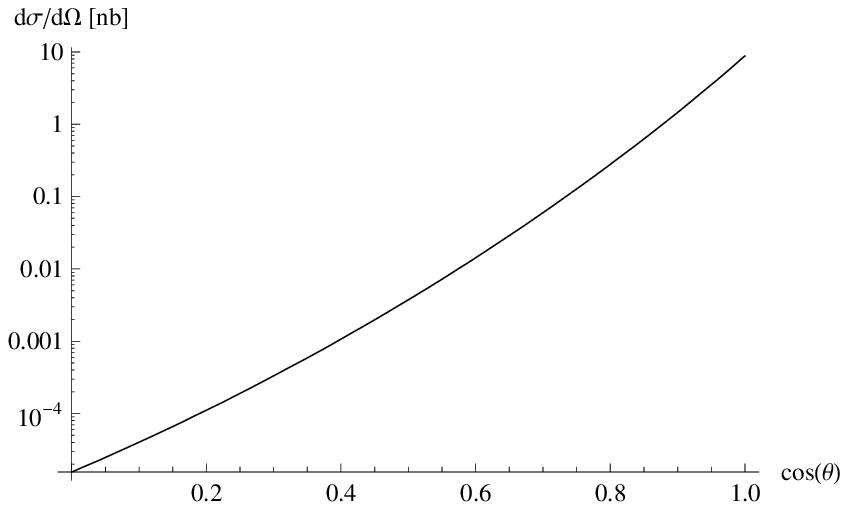,
   height=4.4cm,clip=}}
   \hfill
   \subfigure{\epsfig{file=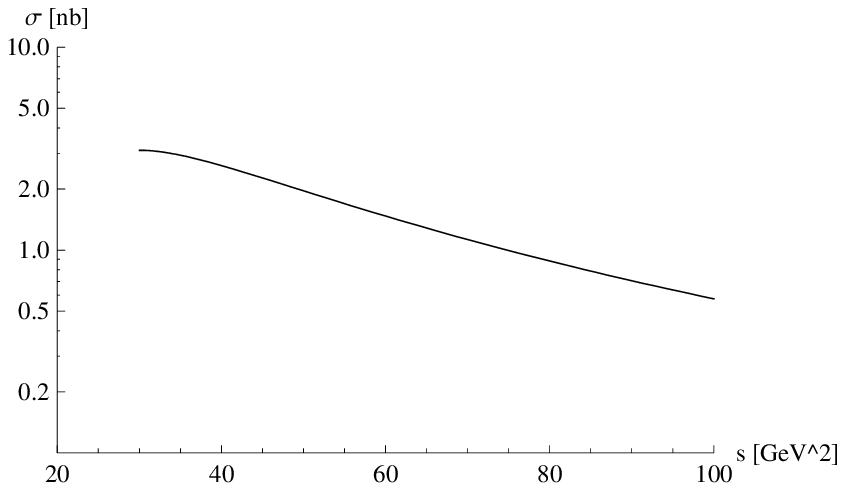,height=4.3cm,clip=}}
   \vspace{-0.3cm}\caption{{\em Left:} The $p  \bar{p} \, \rightarrow
    \, \Lambda_c \bar{\Lambda}_c$ differential cross section $d\sigma
    / d\Omega$
   versus $\cos \theta_{\mathrm{cm}}$ at Mandelstam $s=30\text{ GeV}^2$.
            {\em Right:} The $p  \bar{p} \, \rightarrow
    \, \Lambda_c \bar{\Lambda}_c$ integrated cross section $\sigma$
   versus Mandelstam $s$.}
    \label{Fig_CrossSection}
\end{figure}

For our numerical calculations the transverse size parameters
$a_\Lambda=a_p=1.1$~GeV$^{-1}$ are chosen in such a way that they
provide a value of $\approx 300$~MeV for the mean intrinsic
transverse momentum $<k_\perp^2>^{1/2}$ of a quark inside a
baryon. The wave function normalizations are fixed such that the
probabilities to find the quark-diquark states in a proton or a
$\Lambda_c$ are $0.5$ and $0.9$, respectively. The $x$-dependence
for the resulting transition GPD $H$ at different values of $t$
(i.e. different values of the skewness parameter $\xi$) is shown
in Fig.~\ref{Fig_GPDandFF}. Shown in the same figure is also the
corresponding (scaled) transition form factor $R$ as a function of
$t$. With this form factor the unpolarized differential and
integrated cross sections for  $p \bar{p} \, \rightarrow \,
\Lambda_c \bar{\Lambda}_c$ are easily evaluated. Corresponding
predictions are plotted in Fig.~\ref{Fig_CrossSection}. The
integrated cross section is of the order of nb, i.e. comparable in
magnitude with the result in Ref.~\cite{Kroll:1988cd}. This is
still in the range of high precision experiments. Whether
predictions from more realistic three-quark wave functions will be
comparable to the outcome of this simple quark-diquark model is
presently under investigation~\cite{Goritschnig:2008}.


\vspace{-0.5cm} \acknowledgments{{\noindent A.T.G. acknowledges
the support of the \lq\lq Fonds zur F\"orderung der
wissenschaftlichen Forschung in \"Osterreich\rq\rq (project DK
W1203-N08).\small}}

\end{document}